\documentclass[useAMS, usenatbib, amssymb]{mn2e}
\usepackage{psfig}
\usepackage{graphicx}

\title [Absolute magnitude calibrations for the thin disc stars]
{Luminosity--Colours relations for thin disc main–-sequence stars}

\author[Bilir et al.]
       {S. Bilir,${^1} \thanks{E-mail: sbilir@istanbul.edu.tr}$
        S. Karaali$^{2}$, S. Ak$^{1}$, E. Yaz$^{1}$, A. Cabrera-Lavers$^{3,4}$, 
        K. B. Co\c skuno\u glu$^{1}$
\\
  $^1$Istanbul University Science Faculty, Department of Astronomy and Space 
Sciences, 34119, University-Istanbul, Turkey\\
  $^2$Beykent University, Faculty of Science and Letters, Department of Mathematics  
and Computer, Ayaza\u ga 34396, Istanbul, Turkey\\
  $^3$Instituto de Astrof\'{\i}sica de Canarias, E-38205 La Laguna, Tenerife, Spain\\
  $^4$GTC Project Office, E-38205 La Laguna, Tenerife, Spain\\
}

\date{Accepted 2008 month day.
Received year month day; }

\pagerange{\pageref{firstpage}--\pageref{lastpage}} \pubyear{2008}

\begin{document}

\maketitle

\label{firstpage}

\begin{abstract}
In this study we present the absolute magnitude calibrations of thin disc 
main-sequence stars in the optical ($M_{V}$), and in the near-infrared 
($M_{J}$). Thin disc stars are identified by means of {\em Padova} isochrones, 
and absolute magnitudes for the sample are evaluated via the newly reduced 
{\em Hipparcos} data. The obtained calibrations cover a large range of spectral 
types: from A0 to M4 in the optical and from A0 to M0 in the near-infrared. 
Also, we discuss the of effects binary stars and evolved stars on the absolute 
magnitude calibrations. The usage of these calibrations can be extended to the 
estimation of galactic model parameters for the thin disc individually, 
in order to compare these parameters with the corresponding ones estimated by 
$\chi{^2}_{min}$ statistics (which provides galactic model parameters for 
thin and thick discs, and halo simultaneously) to test any degeneracy between 
them. The calibrations can also be used in other astrophysical researches 
where distance plays an important role in that study.      
\end{abstract}


\section{Introduction}
\label{intro}
The distance of an astronomical object plays an important role in deriving 
intrinsic luminosities of stars, in calculating accurate masses for 
binary system components and in answering questions about galactic structure. 
Particularly, the distributions of three  populations (i.e. thin and thick discs, 
and halo) in the Galaxy can be determined using the distances of the 
concerning stars. For nearby stars, the most appropriate procedure for distance 
determination is the trigonometric parallax. However, this procedure fails 
for distant stars due to large errors in their trigonometric parallaxes. 
In this case, photometric parallax replaces the trigonometric one. This 
alternative requires absolute magnitude determination which needs a 
lot of work, though not as much as the former one. 

The photometric parallax can be evaluated in different ways. Many 
astronomers prefer an absolute magnitude-colour diagram for a star category. 
For example, \citet{Phleps00} separated their star sample into two sub–samples, 
disc and halo, according to their $(r-i)$ colours and they used two absolute 
magnitude diagrams to evaluate their absolute magnitudes. 
\citet*{Karaali03, KBH04} and \citet*{BKG06} separated their star samples into 
three populations, i.e. thin and thick discs, and halo, and they used the absolute 
magnitude–-colour diagrams of three globular clusters with the same metallicity 
of the corresponding population. \citet*{Bilir05} calibrated the $M_{g}$ absolute 
magnitudes of late type disc dwarfs with $(g-r)$ and $(r-i)$ colours in 
{\em SDSS} system. The works of \citet{Siegel02} and \citet{Juric08} are examples 
for distance estimation based on photometric parallax. Another procedure for 
absolute magnitude determination is based on the use of colours and individual 
UV--excesses of stars relative to a standard absolute magnitude–-colour diagram, 
such as Hyades \citep*{Laird88,Karaali03,KBT05}. In this procedure, one does not 
need to separate the stars into different population types. Additionally, individual 
UV--excess for each star results in more precise absolute magnitudes relative 
to the procedure where a single colour magnitude diagram is used for all stars in a 
population.

Despite the extensive applications of the aforementioned procedures in the previous 
paragraph, additional constraints can be considered for obtaining a more reliable one. 
We applied three limitations, i.e. age, metallicity and surface gravity, 
to a star sample and calibrate the absolute magnitude of thin disc stars as a 
function of two colours. All these data were provided from {\em Padova} database of 
stellar evolutionary tracks and isochrones \citep{Marigoetal08} by using a web 
interface\footnote{http://stev.oapd.inaf.it/$\sim$ lgirardi/cgi-bin/cmd}. For the 
thin disc stars we adopted a range of age of $0\leq t\leq10$ Gyr, 
and the metal abundance is assumed to be $0.01\leq z \leq 0.03$ (solar metal 
abundance $z_{\odot}=0.019$), corresponding to the metallicity interval 
$-0.30\leq[M/H]\leq0.20$ dex. Finally, evolved stars were excluded from the 
sample by imposing a third constraint, i.e. surface gravity $\log g>4$. Thus, 
we should supply a homogeneous thin disc main–-sequence sample by eliminating all 
thick disc and halo stars, as well as evolved (white dwarfs, sub–-giants and giants) 
thin disc stars.                 

\section{Data and reductions}

{\em BVI} photometric data and $\pi$ parallaxes with relative errors 
($\sigma_{\pi}/\pi)\leq 0.05$ for 11644 stars were taken from the newly reduced 
{\em Hipparcos} data \citep{vanLeeuwen2007}, whereas the {\em 2MASS} 
(Two Micron All Sky Survey) near-infrared photometric data for the same stars 
were extracted of the Point-Source Catalogue and Atlas \citep{Cu03}. The 
{\em 2MASS} photometric system comprises Johnson's $J$ (1.25 $\mu$m) and $H$ 
(1.65 $\mu$m) bands with the addition of $K_{s}$ (2.17 $\mu$m) band, which is 
bluer than Johnson's $K$-band \citep{Skrutskie06}. The $E(B-V)$ colour–-excesses 
were individually evaluated for each sample star making use of the maps of 
\citet*{Schlegel98}, and this was reduced to a value corresponding to the distance 
of the star by means of the equations of \citet{Bahcall80}. The $E(B-V)$ 
iso--colour excess contours within 100 pc of solar neighbourhood are given in Fig. 1. 
The distribution of $E(B-V)$ for the star sample for two distance intervals, 
$0<d\leq40$ pc and $40<d\leq70$ pc, {\bf have a} peak at $E(B-V)=0.0041$ and $E(B-V)=0.0098$ mag 
which are rather close to the ones of \citet*{Holmberg07}. \citet{Holmberg07} 
evaluated the colour excesses, $E(B-V)=0.0034$ and $E(B-V)=0.0065$ mag, for the 
same distance intervals by using Str\"omgren photometry, respectively. The 
transformation of $E(b-y)$ reddening to the $E(B-V)$ one is carried out by the 
equation $E(B-V)=1.35\times E(b-y)$. The excellent agreement between both sets 
of reddening data confirms the reduction equations of \citet{Bahcall80} and the 
accuracy of the colour–-excesses used in our work.

Thus, colours and magnitudes from the sample stars are de–-reddened by using the 
following two sets of equations, one for the {\em BVI} data (Eq. 1) and one for 
the {\em 2MASS} (Eq. 2) photometry: 

\begin{eqnarray}
V_{0}=V-3.1\times E(B-V),\nonumber\\
(B-V)_{0}=(B-V)-E(B-V),\\
(V-I)_{0}=(V-I)-1.25\times E(B-V).\nonumber
\end{eqnarray} 

\begin{eqnarray}
J_{0}=J-0.887\times E(B-V),\nonumber\\
(J-H)_{0}=(J-H)-0.322\times E(B-V),\\
(H-K_{s})_{0}=(H-K_{s})-0.182\times E(B-V).\nonumber
\end{eqnarray} 
\citep[see also,][]{Biliretal06, Aketal07}.

The absolute magnitude of each star is evaluated by the combination of its 
de–-reddened apparent magnitude and its distance estimated using its trigonometric 
parallax. The corresponding propagated errors were calculated as 
$\sigma M=2.17(\sigma_{\pi}/\pi)+\sigma m$, where ($\sigma_{\pi}/\pi$) and 
$\sigma m$ are the relative parallax error and the error of the apparent 
magnitude in the relevant photometric system, respectively. 

\begin{figure}
\center
\resizebox{80mm}{80mm}{\includegraphics*{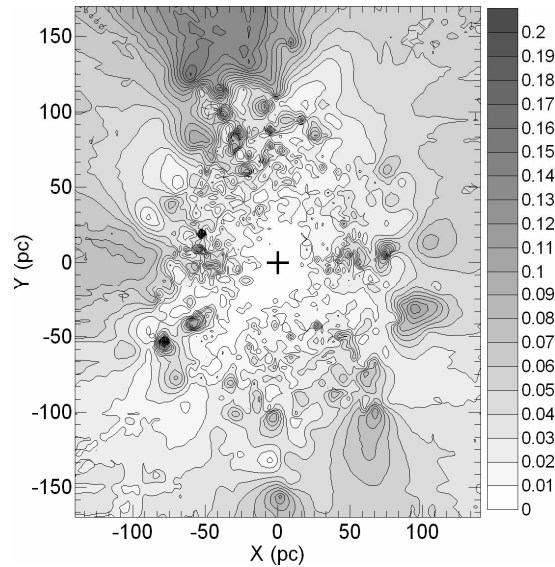}}
\caption{The $E(B-V)$ iso-colour-excess contours within 100 pc of the solar 
neighbourhood. Here, $X$ and $Y$ are heliocentric galactic coordinates 
oriented towards the galactic centre and galactic rotation, respectively.}
\end{figure} 
 
\begin{figure}
\center
\includegraphics[scale=0.4, angle=0]{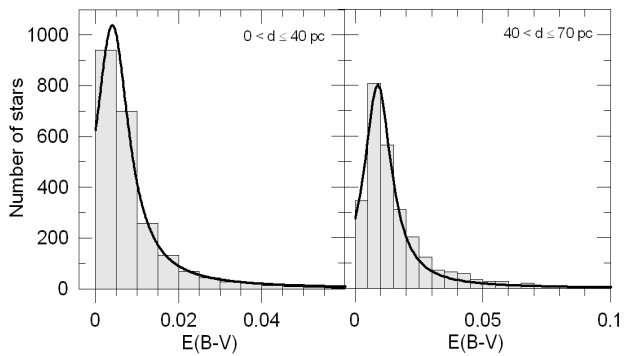}
\caption{Distribution of the $E(B-V)$ colour--excess for the star sample for 
two distance intervals $0<d\leq40$ and $40<d\leq70$ pc. The $E(B-V)$ 
colour-excesses corresponding to the these distributions are in good agreement 
with the ones of \citet{Holmberg07}.} 
\end{figure} 
 
\section{The procedure and absolute magnitude calibration}

The procedure consists of the calibration of the absolute magnitudes for thin 
disc main–-sequence stars with two colours: one sensitive to early--type (hot) 
and another sensitive to late--type (cool) stars, providing a large 
range of absolute magnitudes for the thin disc population. We calibrated 
$M_{V}$ absolute magnitudes with $(B-V)_{0}$ and $(V-I)_{0}$, for the {\em BVI} 
photometry and $M_{J}$ absolute magnitudes with $(J-H)_{0}$ and $(H-K_{s})_{0}$ 
for the {\em 2MASS} data. As it was stated in Section \ref{intro}, we limited 
the metallicity, the age, and the surface gravity with the following constraints: 
$-0.30\leq[M/H]\leq0.20$ dex, $0\leq t \leq 10$ Gyr, and $\log g>4$. By doing this we avoid 
any contamination due to thick disc and halo stars, as well as from evolved thin 
disc stars. Hence, our calibrations should provide precise absolute magnitudes 
for thin disc main--sequence stars.

\subsection{Absolute magnitude calibration for {\em BVI} photometry}

We applied a series of limitations in the absolute magnitude calibration 
for  {\em BVI} photometry. First of all, we limited our star sample with absolute 
magnitudes $0<M_{V}<12$. This limitation reduced the original star sample from 
11644 to 10654. Then, we applied the procedure described in Section 2 to the 
{\em Padova} isochrones with metallicity and age limitations mentioned above 
\citep{Marigoetal08} which separates the thin disc stars with different luminosity 
classes (dwarf and evolved stars) from the sample of 10654 stars. At the third 
step, we applied the constraint for being a main--sequence star, i.e. $\log g>4$. 
Thus, the sample reduced to 6117 stars. Fig. 3 shows the colour--absolute magnitude 
diagram for the original sample (11644 stars) and the upper and lower envelopes of 
the final sample, i.e. thin disc main-sequence stars. Then we adopted the $M_{V}$ 
calibration as follows and evaluated the coefficients by the least–-squares 
method, for the final sample:

\begin{eqnarray}
M_{V} =  a_{1}(B-V)^{2}_{0}+ b_{1}(V-I)^{2}_{0}+ c_{1}(B-V)_{0}(V-I)_{0}\\
 +d_{1}(B-V)_{0}+e_{1}(V-I)_{0}+f_{1}.\nonumber
\end{eqnarray}

The numerical values of the coefficients and their errors, as well as the corresponding 
standard deviations and the squared correlation coefficients are all given in 
Table 1. The calibration described by Eq. 3 covers a large range of thin disc 
main–-sequence stars, i.e. $-0.15<(B-V)_{0}<1.60$, $-0.15<(V-I)_{0}<2.90$, 
and $0<M_{V}<12$, that corresponds to the spectral types A0-–M4. 

\begin{figure}
\center
\includegraphics[scale=0.35, angle=0]{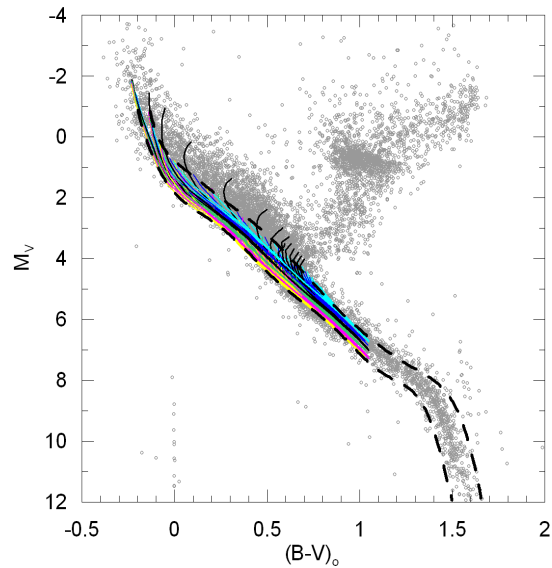}
\caption{$M_{V}/(B-V)_{0}$ colour-absolute magnitude diagram for the original 
sample. The upper and lower envelopes (the dashed lines) show the final sample, 
i.e. thin disc main--sequence stars. The thin curves correspond to {\em Padova} 
isochrones.}
\end{figure} 

\subsection{Absolute magnitude calibration for {\em 2MASS} photometry}
The procedure described in Section 2 and the limitations applied in Section 3.1 
produced 4449 main--sequence stars with {\em 2MASS} photometric data, 93 per cent 
of the best quality following the survey criteria (labeled in the catalogue as 
AAA). Here, the star sample is limited with absolute magnitude $0<M_{J}<6$. 
The $M_{J}/(J-H)_{0}$ colour–-magnitude diagram for all stars taken from newly 
reduced {\em Hipparcos} catalogue (11644 stars) and the upper and lower envelopes 
for the final sample, i.e. thin disc main-sequence stars, are given in 
Fig. 4. We adopted an absolute magnitude calibration for the {\em 2MASS} data 
similar to the {\em BVI} ones as follows:

\begin{eqnarray}
M_{J}=a_{2}(J-H)^{2}_{0}+ b_{2}(H-K_{s})^{2}_{0}+ c_{2}(J-H)_{0}(H-K_{s})_{0}\\
+d_{2}(J-H)_{0}+e_{2}(H-K_{s})_{0}+f_{2}.\nonumber
\end{eqnarray}
The numerical values of the coefficients and their errors, the corresponding 
standard deviations, and the squared correlation coefficients are given in 
Table 1. The calibration given in Eq. 4 covers a large range of thin disc 
main–-sequence stars, as the one in Eq. 3, i.e. $-0.16<(J-H)_{0}<0.70$, 
$-0.07<(H-K_{s})_{0}<0.26$, and $0<M_{J}<6$ corresponding to the spectral 
types A0-–M0.

We plotted the errors of the observed colours against the intrinsic colours in 
Fig. 5 in order to test the accuracy of the observed colours. The lower 
uncertainties belong to $(B-V)_{0}$ and $(V-I)_{0}$ colours, whereas those for 
$(J-H)_{0}$ and $(H-K_{s})_{0}$ are larger. The mean observational errors 
are about 0.01 ($\sigma=\pm0.02$) and 0.04 ($\sigma=\pm0.04$) mag in the optical 
and near-infrared colours, respectively. This is not surprising, because {\em 2MASS} 
magnitudes were obtained from single-epoch observations, whereas optical 
magnitudes have been observed more than once. The mean errors introduce typically 
$\pm$0.12 and $\pm$0.14 mag uncertainties in $M_{V}$ and $M_{J}$, respectively. 

\begin{figure}
\center
\includegraphics[scale=0.35, angle=0]{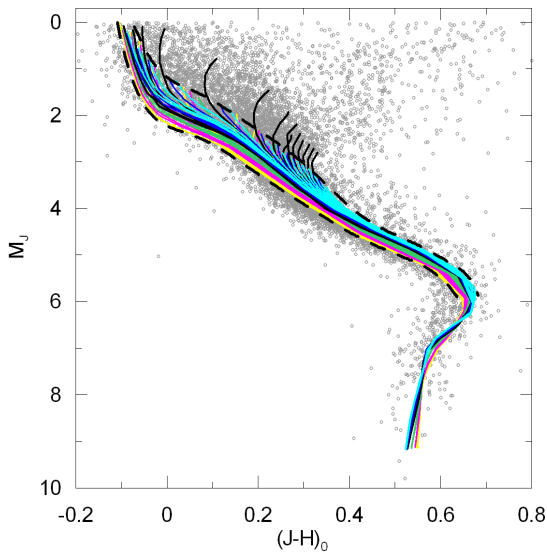}
\caption{$M_{J}/(J-H)_{0}$ colour-absolute magnitude diagram for the original 
sample. The upper and lower envelopes (the dashed lines) show the final sample, 
i.e. thin disc main--sequence stars. The thin curves correspond to {\em Padova} 
isochrones.}
\end{figure}

\begin{table*}
\setlength{\tabcolsep}{2pt} 
\center 
\caption{Coefficients and their standard errors for the Eqs. 3 ($i$=1) and 4 ($i$=2). 
$R^{2}$ and $s$ denotes the squared correlation coefficient and the standard 
deviation, respectively.}
\begin{tabular}{ccccccccc}
\hline
Eq. &  $a_{i}$ &  $b_{i}$  &  $c_{i}$ & $d_{i}$ & $e_{i}$ &  $f_{i}$ & $R^{2}$  &    $s$ \\
\hline
3 & -4.003~($\pm$0.532) &  0.837~($\pm$0.192) & 2.633~($\pm$0.671) & 11.796~($\pm$0.359) & -5.691~($\pm$0.323) & 1.122~($\pm$0.011) & 0.98  & 0.27\\
4 & -1.732~($\pm$0.109) & -7.734~($\pm$0.846) & 1.084~($\pm$0.449) & ~7.509~($\pm$0.048) &  2.208~($\pm$0.166) & 1.305~($\pm$0.008) & 0.98  & 0.19\\
\hline
\end{tabular} 
\end{table*}

\begin{figure}
\center
\includegraphics[scale=0.4, angle=0]{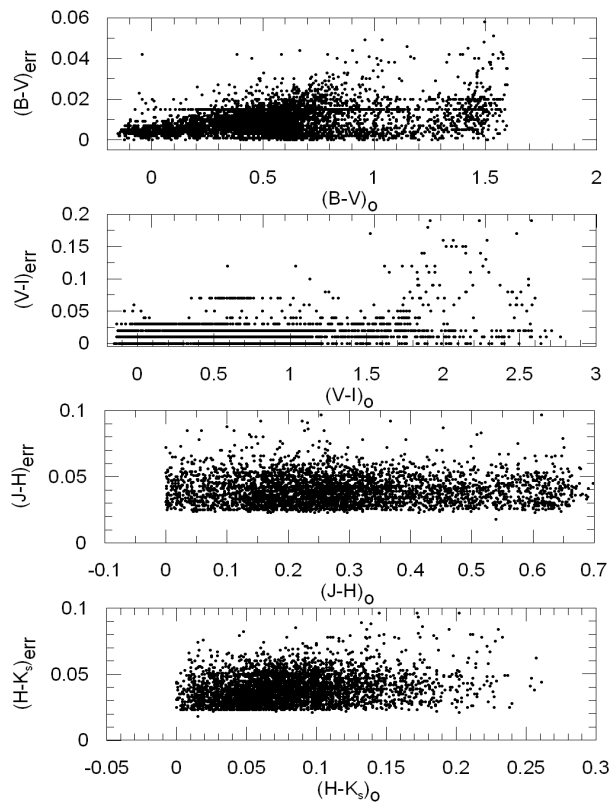}
\caption{Colour errors for the {\em BVI} and {\em 2MASS} photometric data.}
\end{figure} 

Unfortunately, random errors, presumably symmetric on the measured parallaxes, 
do not provide symmetric uncertainties on the computed distances. Therefore, a 
measured trigonometric parallax is very likely to be larger than the true 
parallax. The problem has already been noticed and studied by \cite{LK73}. 
Assuming a uniform space distribution of stars and a Gaussian distribution of 
observed parallaxes about a true parallax, \cite{LK73} revealed that there 
is a systematic error in the computed distances which depends only upon the 
ratio ($\sigma_{\pi}/\pi$), where $\pi$ is the observed parallax. \cite{J01}  
showed that only the studies which are careful enough to use parallaxes with 
$(\sigma_{\pi}/\pi)<0.1$ could be excused as the bias would be negligible. 
This is the case in our work, where ($\sigma_{\pi}/\pi)\leq 0.05$. Actually, the 
Lutz-Kelker correction in absolute magnitude, taken from \cite{LK73}, is less 
than 0.03 mag. Hence, we omitted the mentioned bias in our study. Although no 
absolute magnitude calibration based on trigonometric parallaxes is present for 
the $M_{g}$ absolute magnitude in {\em SDSS} system, one can use our recent 
transformation equations \citep{Biliretal08}.   

\subsection{Comparison of the estimated absolute magnitudes with the 
trigonometric parallaxes and synthetic photometry}

We compared the  absolute magnitudes estimated in this work with two sets of 
absolute magnitudes, one evaluated by means of the trigonometric parallaxes 
taken from the newly reduced {\em Hipparcos} catalogue \citep{vanLeeuwen2007} 
and one taken from the stellar spectral flux library of \citet{Pickles98}. 
Figs. 6 and 7 show the one–-to-one correspondence of the absolute magnitudes 
$M_{V_{c}}$ and $M_{J_{c}}$, respectively, estimated in this work and the 
corresponding ones evaluated from the newly reduced {\em Hipparcos} data, i.e. 
$M_{V_{Hip}}$ and $M_{J_{Hip}}$. 

\citet{Pickles98} offers the synthetic colours, $M_{bol}$ bolometric absolute 
magnitudes and {\em BC} bolometric corrections for 131 stars with a large range of 
spectral type, O5–-M6, and  different luminosity-–classes. The optical data 
are on the same scale of {\em UBVRI} photometry. Hence, it was easy to evaluate 
the $M_{V_{c}}$ absolute magnitudes, by placing $(B-V)_{0}$ and $(V-I)_{0}$ colours 
into Eq. 3. However, the infrared colours and magnitudes scale differently than the 
{\em 2MASS} data. Hence, we used the normalized equations of \citet{Carpenter01} 
to reduce Pickles' (1998) infrared data to {\em 2MASS} colours and 
magnitudes. Then, we evaluated the $M_{J_{c}}$ absolute magnitudes by placing 
the reduced $(J-H)_{0}$ and $(H-K_{s})_{0}$ colours into Eq. 4. The original data of 
\citet{Pickles98} and the reduced ones according to normalized equations of 
\citet{Carpenter01} are given in Tables 2 and 3, respectively. Finally, 
we evaluated the $M_{V_{Pic}}$ and $M_{J_{Pic}}$ absolute magnitudes from the 
data in Table 3 and we compared them with the $M_{V_{c}}$ and $M_{J_{c}}$ 
absolute magnitudes, respectively (Figs. 8 and 9). There is a one--to--one 
correspondence in these figures as well. The slight declination at the faint 
end of absolute magnitudes in Fig. 9 is probably due to the different scales 
between the Pickles' (1998) and {\em 2MASS}. We should add that the mentioned 
comparison has been carried out only for stars of solar metallicity. 

\begin{figure}
\center
\includegraphics[scale=0.35, angle=0]{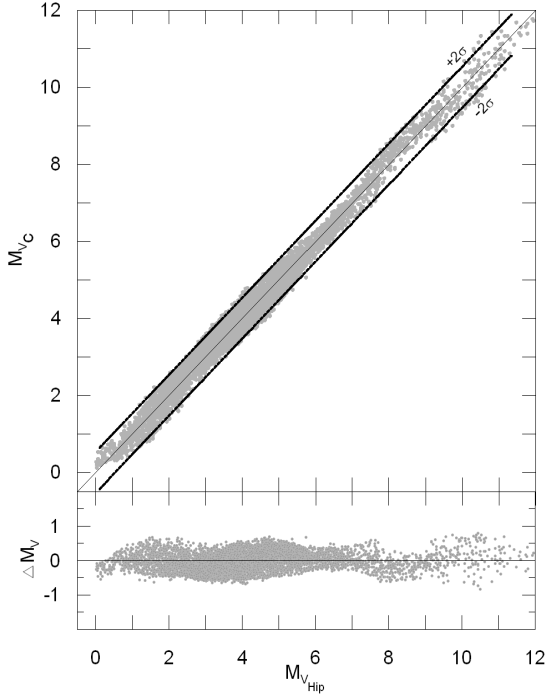}
\caption{Absolute magnitudes, estimated by Eq. 3, versus optical absolute magnitudes 
calculated from newly reduced {\em Hipparcos} data (upper panel) and variation of the differences 
between two sets of absolute magnitudes (lower panel). All calibration stars in the 
figure are located in the prediction limit of $2\sigma$.}
\end{figure}

\begin{figure}
\center
\includegraphics[scale=0.33, angle=0]{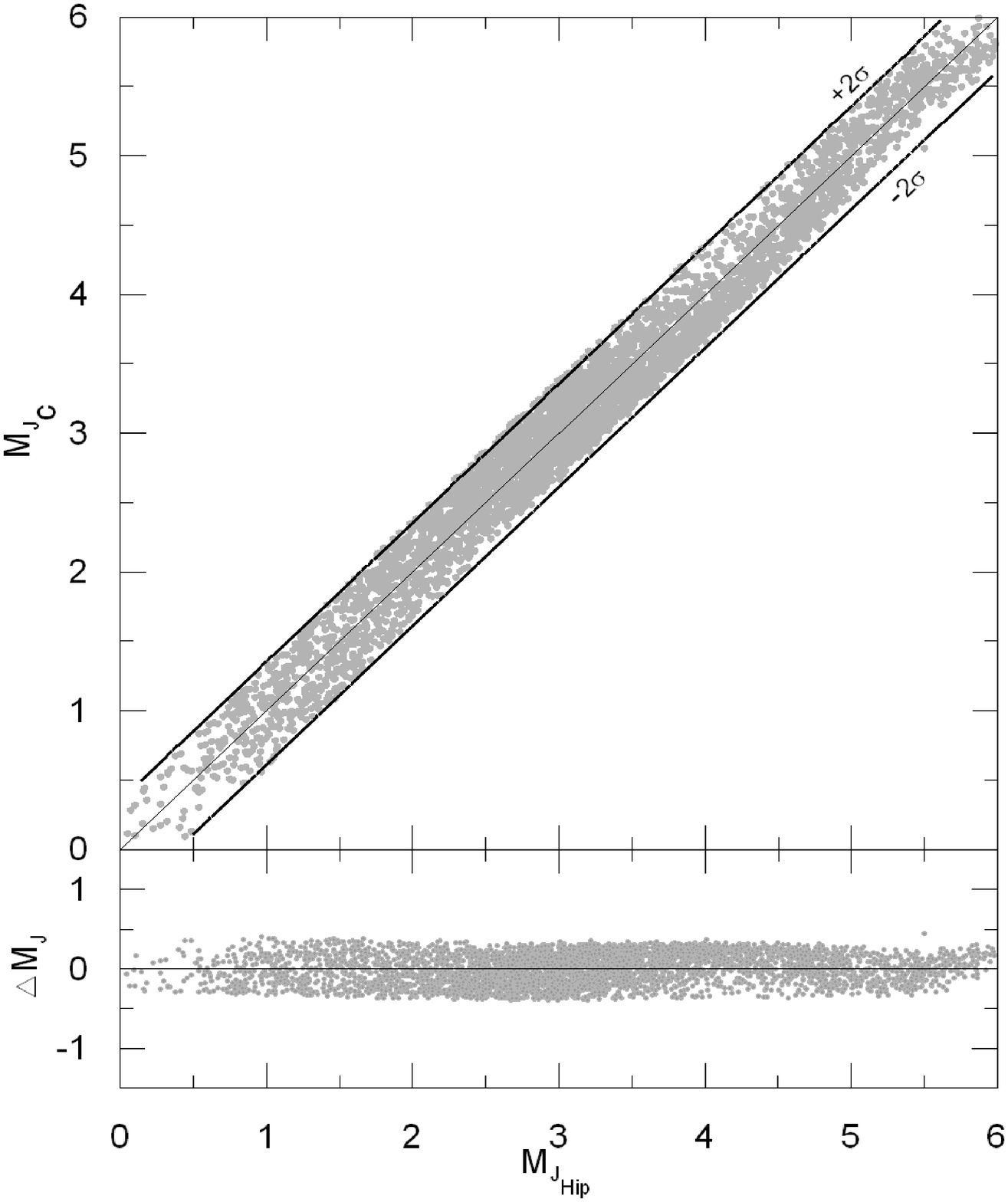}
\caption{Absolute magnitudes, estimated by Eq. 4, versus near-infrared absolute 
magnitudes calculated from newly reduced {\em Hipparcos} data (upper panel) 
and variation of the differences between two sets of absolute magnitudes (lower panel). 
All calibration stars in the figure are located in the prediction limit of $2\sigma$.}
\end{figure}

\begin{figure}
\center
\includegraphics[scale=0.34, angle=0]{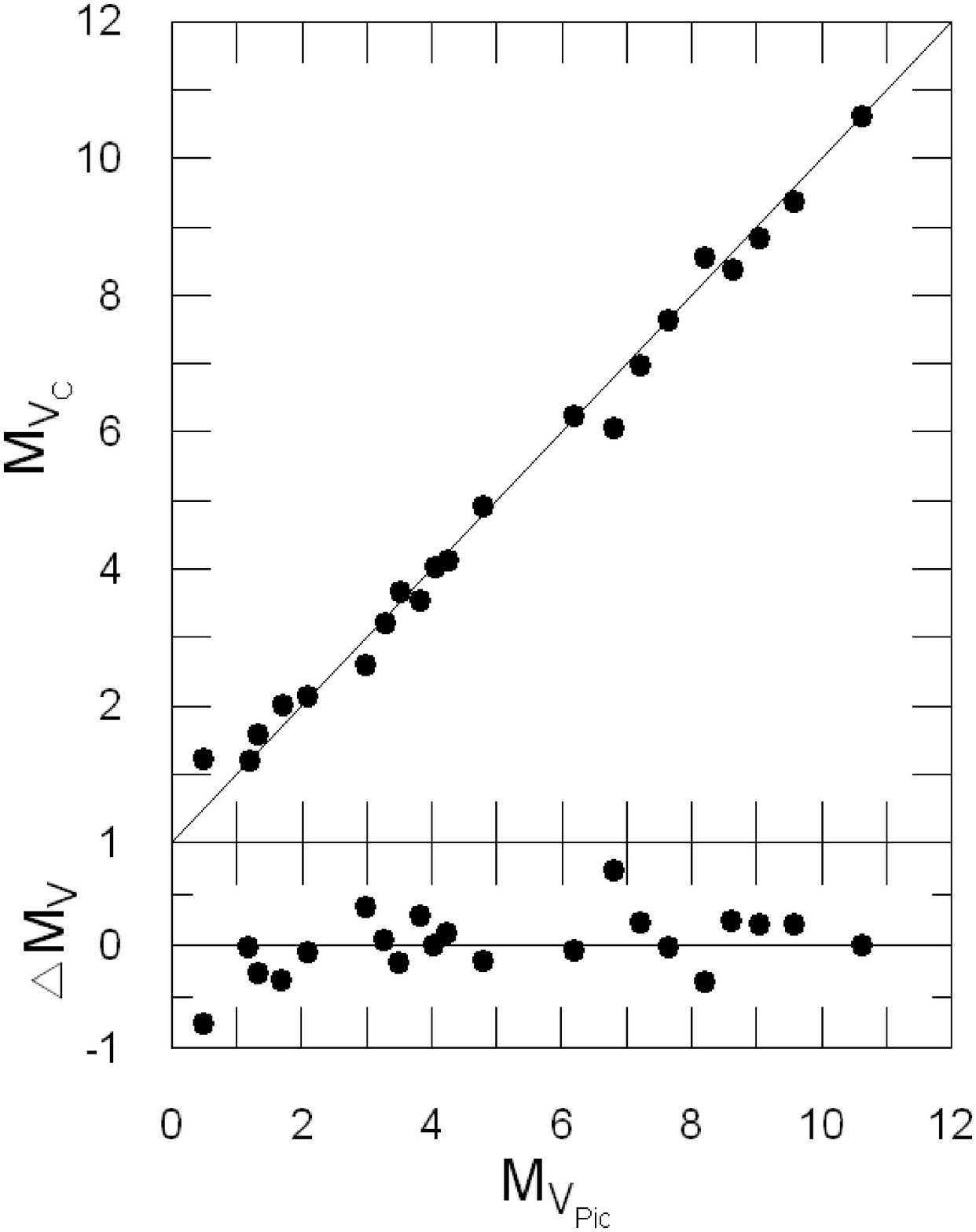}
\caption{Absolute magnitudes, estimated by Eq. 3, versus optical absolute magnitudes 
calculated from Pickles' data (upper panel) and variation of the differences between 
two sets of absolute magnitudes (lower panel).}
\end{figure}

\begin{figure}
\center
\includegraphics[scale=0.30, angle=0]{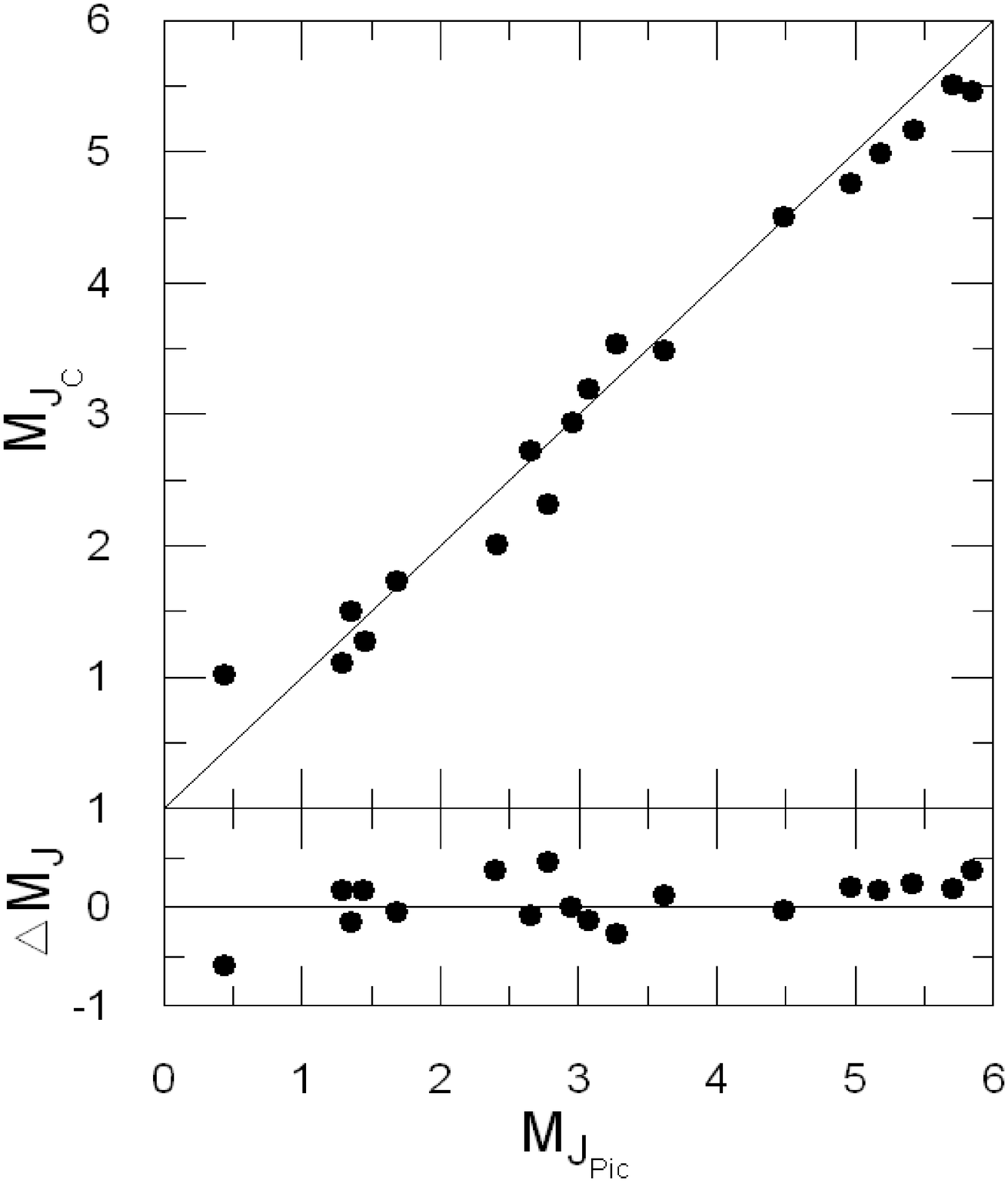}
\caption{Absolute magnitudes, estimated by Eq. 4, versus near-infrared absolute 
magnitudes calculated from Pickles' data (upper panel) and variation of the 
differences between two sets of absolute magnitudes (lower panel).}
\end{figure}

\begin{table}
\setlength{\tabcolsep}{1.3pt}
\center \caption{Original photometric data of \citet{Pickles98}.}
\begin{tabular}{|cccccccc}
\hline
 Sp. Type & $(B-V)$ & $(V-I)$ & $(R-I)$ & $(J-H)$ &   $(H-K)$ &  $M_{V_{Pic}}$ &  $M_{K_{Pic}}$ \\
\hline
       A0V &      0.015 &      0.011 &      0.023 &      0.000 &      0.000 &       0.48 &       0.49 \\
       A2V &      0.029 &      0.049 &      0.043 &      0.010 &      0.010 &       1.18 &       1.32 \\
       A3V &      0.089 &      0.102 &      0.065 &      0.030 &      0.020 &       1.33 &       1.45 \\
       A5V &      0.153 &      0.156 &      0.103 &      0.060 &      0.020 &       1.69 &       1.33 \\
       A7V &      0.202 &      0.241 &      0.131 &      0.090 &      0.030 &       2.09 &       1.62 \\
       F0V &      0.303 &      0.378 &      0.203 &      0.130 &      0.030 &       2.98 &       2.30 \\
       F2V &      0.395 &      0.457 &      0.246 &      0.170 &      0.040 &       3.27 &       2.63 \\
       F5V &      0.458 &      0.496 &      0.255 &      0.230 &      0.040 &       3.50 &       2.44 \\
       F6V &      0.469 &      0.562 &      0.292 &      0.260 &      0.050 &       3.83 &       2.70 \\
       F8V &      0.542 &      0.615 &      0.312 &      0.300 &      0.040 &       4.04 &       2.79 \\
       G0V &      0.571 &      0.671 &      0.351 &      0.350 &      0.050 &       4.24 &       2.94 \\
       G5V &      0.686 &      0.735 &      0.372 &      0.340 &      0.070 &       4.78 &       3.27 \\
       K2V &      0.924 &      0.968 &      0.448 &      0.500 &      0.090 &       6.19 &       3.96 \\
       K3V &      0.930 &      1.109 &      0.513 &      0.540 &      0.100 &       6.80 &       4.40 \\
       K4V &      1.085 &      1.232 &      0.570 &      0.580 &      0.110 &       7.21 &       4.56 \\
       K5V &      1.205 &      1.361 &      0.610 &      0.610 &      0.110 &       7.64 &       4.77 \\
       K7V &      1.368 &      1.578 &      0.750 &      0.660 &      0.150 &       8.21 &       5.11 \\
       M0V &      1.321 &      1.709 &      0.847 &      0.670 &      0.170 &       8.62 &       4.94 \\
       M1V &      1.375 &      1.874 &      0.993 &      0.660 &      0.280 &       9.05 &       5.16 \\
       M2V &      1.436 &      2.020 &      1.061 &      0.660 &      0.200 &       9.58 &       5.44 \\
       M3V &      1.515 &      2.436 &      1.362 &      0.640 &      0.230 &      10.63 &       5.96 \\
       M4V &      1.594 &      2.781 &      1.565 &      0.620 &      0.270 &      11.54 &       6.24 \\
  
\hline
\end{tabular} 
\end{table}

\begin{table}
\setlength{\tabcolsep}{2pt}
\center 
\caption{Data reduced by normalized equations of \citet{Carpenter01}. The 
columns give: (1) Spectral type, (2) $M_{V_{Pic}}$ absolute magnitude by Pickles 
(the same as in Table 2), (3) $M_{V_{c}}$ absolute magnitude evaluated by Eq. 3, 
(4) and (5) and (6) reduced $(J-H)$, $(H-K_{s})$ and $M_{J}$ data ({\em 2MASS}), 
and (7) $M_{J}$ absolute magnitude evaluated by Eq. 4.}    
\begin{tabular}{ccccccc}
\hline
(1) & (2) & (3) & (4) & (5) & (6) & (7) \\
\hline
Sp. Type   & $M_{V_{Pic}}$  & $M_{V_{c}}$ & $(J-H)$ & $(H-K_{s})$ & $M_{J}$ & $M_{J_{c}}$\\
\hline
       A0V &       0.48 &       1.24 &     -0.045 &      0.028 &       0.44 &       1.02 \\
       A2V &       1.18 &       1.19 &     -0.035 &      0.038 &       1.28 &       1.11 \\
       A3V &       1.33 &       1.59 &     -0.016 &      0.048 &       1.44 &       1.27 \\
       A5V &       1.69 &       2.03 &      0.014 &      0.048 &       1.35 &       1.50 \\
       A7V &       2.09 &       2.15 &      0.043 &      0.058 &       1.68 &       1.73 \\
       F0V &       2.98 &       2.60 &      0.082 &      0.058 &       2.40 &       2.02 \\
       F2V &       3.27 &       3.21 &      0.122 &      0.068 &       2.78 &       2.32 \\
       F5V &       3.50 &       3.67 &      0.180 &      0.068 &       2.65 &       2.73 \\
       F6V &       3.83 &       3.53 &      0.210 &      0.078 &       2.95 &       2.95 \\
       F8V &       4.04 &       4.03 &      0.249 &      0.068 &       3.07 &       3.20 \\
       G0V &       4.24 &       4.12 &      0.298 &      0.078 &       3.27 &       3.54 \\
       G5V &       4.78 &       4.93 &      0.288 &      0.098 &       3.61 &       3.50 \\
       K2V &       6.19 &       6.24 &      0.445 &      0.118 &       4.48 &       4.51 \\
       K3V &       6.80 &       6.07 &      0.484 &      0.128 &       4.97 &       4.76 \\
       K4V &       7.21 &       6.99 &      0.523 &      0.138 &       5.18 &       4.99 \\
       K5V &       7.64 &       7.65 &      0.553 &      0.138 &       5.42 &       5.17 \\
       K7V &       8.21 &       8.56 &      0.602 &      0.177 &       5.84 &       5.46 \\
       M0V &       8.62 &       8.38 &      0.612 &      0.197 &       5.70 &       5.52 \\
       M1V &       9.05 &       8.83 &      0.602 &      0.307 &       6.02 &        --- \\
       M2V &       9.58 &       9.37 &      0.602 &      0.227 &       6.22 &        --- \\
       M3V &      10.63 &      10.63 &      0.582 &      0.257 &       6.75 &        --- \\
       M4V &      11.54 &      12.08 &      0.563 &      0.297 &       7.05 &        --- \\
\hline
\end{tabular} 
\end{table}

\section{Discussion}
We present two equations (Eqs. 3 and 4) derived from newly reduced {\em Hipparcos} 
data \citep{vanLeeuwen2007} with the aim of applying these formulae to relatively 
distant stars whose distances are either newly reduced not accurately known or not known 
at all. To do this, one needs to obtain a sample of thin disc stars and replace 
the $(B-V)_{0}$ and $(V-I)_{0}$ colours with Eq. 3 or $(J-H)_{0}$ and $(H-K_{s})_{0}$ 
colours with Eq. 4, depending on the preferred photometry. This procedure 
can supply the following contributions to the estimation of galactic model parameters:
\begin{itemize}
\item
One can evaluate the space densities in the solar neighbourhood ($r<400$ pc -- 
we assume all these stars belong to the thin disc) by using the {\em 2MASS} 
data where the {\em SDSS} magnitudes are saturated; and combine them with space 
densities at larger distances evaluated by {\em SDSS} data. Thus, obtaining a 
continuous space density function from the Sun to large distances. This approach 
provides accurate galactic model parameters for the galactic components (thin 
and thick discs, and halo).
\item
This procedure provides individually estimated galactic model parameters for the 
thin disc. Hence, one can compare these parameters with the ones estimated by 
$\chi{^2}_{min}$ statistics which provide galactic model parameters simultaneously 
for thin and thick discs, and halo; and we can test any possible degeneracy between 
different galactic model parameters. We should emphasize that the mentioned 
degeneracy is a serious problem for the galactic model parameters. Thus, we hope 
to contribute a little to this problem which is suffered by the galactic model 
researchers.
\end{itemize}

The absolute magnitude calibrations for the thin disc main-sequence stars 
with two colours, one sensitive to early type (hot) and another 
sensitive to late type (cool) stars, can also be used in other astrophysical 
researches, apart from the galactic model parameters estimation. Since 
absolute magnitude supplies the distance of a star, which plays an important 
role in the investigation of many properties of that star.   

However, there are two significant issues which need to be considered on this 
subject, i.e. binary stars and evolved stars.

\subsection {Binary stars}
A high fraction of stars are in fact binary systems and being a binary system 
makes stars appear brighter and redder than they normally are. Different fractional 
values (defined as $f$) can be found in the literature. For example, using the data 
in the Gliese catalogue of nearby stars, \citet{Brosche64} found a value of $f=0.4$ 
for his simple model for the resolution criterion. On considering the local (within a 
distance of 10 pc) binary fraction, \citet{Reid91} concluded that the proportion of 
binaries among ``stars'' is consistent with a value ranging from 30 to 50 per cent. 
When all systems in question are binary stars, i.e. $f=1$, \citet {Kroupaetal91} found 
that a single mass function provides the best representation of a single luminosity 
function. However, a smaller value can not be discarded with high confidence. 
\citet{Halbwachs86} used all available data on binary systems and concluded that the 
proportion of single stars among all stellar systems is at most 23 per cent when 
spectroscopic binaries are taken into account. An extensive long--term radial 
velocity study of the Hyades cluster reveals that at least 30 per cent of the cluster 
stars are spectroscopic binaries and that essentially all stars brighter than the Hyades 
main--sequence stars are in fact binary systems \citep{Griffinetal88}.

The effect of binary stars were discussed in \citet[][hereafter KTG90 and 
KTG93, respectively]{Kroupaetal90, Kroupaetal93} extensively as well as other 
effects such as metallicity, age, distance etc. KTG93 adopt the binary fraction 
$f\sim0.6-0.7$ as a reasonable value. They give the mentioned combined effects as 
``cosmic scatter'' as a function of $(V-I)$ colour in the range $0.5<(V-I)<4.5$. 
These authors estimate the scatter belonging to binaries alone as  $\sigma=0.27$ mag, 
if a fraction $f=0.8$ of all stars are unresolved binary systems. We adopted a simple 
but reasonable procedure, explained in the following paragraphs, to reveal the binarism 
effect in our work and to compare with the ones appeared in the studies cited above.

We separated the $(B-V)_{0}$ and $(J-H)_{0}$ colours into small bins and omitted a 
fraction of bright stars, each time, in these intervals. Then, we re--calibrated the 
absolute magnitudes $M_{V}$ and $M_{J}$ as a function of colours for the remaining 
stars. The fractions of binaries range from 0 to 80 per cent in steps of 10 per cent. 
Each time the locus of the stars on the colour--magnitude diagram moved to fainter 
absolute magnitudes. The mean of the differences between the absolute magnitudes 
estimated by the calibration in Section 3 and by these loci is adopted as the 
scatter due to the binaries in question (Table 4 and Fig. 10). Our procedure is 
based on the fact that binarism make stars too bright and too red. We calibrated 
the absolute magnitude scatter in $f$--absolute magnitude diagram, $\Delta M_{V}$ 
and $\Delta M_{J}$, linearly (Fig. 11) as follows:

\begin{equation}
\Delta M_{V}=0.411 \times f+0.009,
\end{equation}

\begin{equation}
\Delta M_{J}=0.266\times f+0.014.
\end{equation}

\begin{table}
\setlength{\tabcolsep}{3.5pt}
\center \caption{Scatter in absolute magnitude, $\Delta M_{V}$ and 
$\Delta M_{J}$, as a function of binary fraction $f$.}
\begin{tabular}{ccc}
\hline
$f$ & $\Delta M_{V}$ & $\Delta M_{J}$ \\
\hline
       0.0&      0.000 &      0.000 \\
       0.1&      0.057 &      0.044 \\
       0.2&      0.102 &      0.077 \\
       0.3&      0.136 &      0.101 \\
       0.4&      0.172 &      0.124 \\
       0.5&      0.208 &      0.148 \\
       0.6&      0.248 &      0.171 \\
       0.7&      0.294 &      0.196 \\
       0.8&      0.348 &      0.226 \\  
\hline
\end{tabular} 
\end{table}

\begin{figure}
\center
\includegraphics[scale=0.37, angle=0]{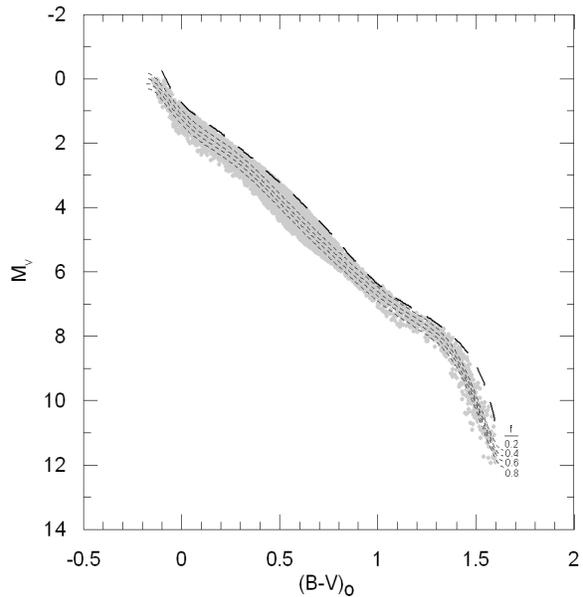}
\caption{Lines indicating the fraction of binary stars in the $M_{V}/(B-V)_{0}$ 
colour--magnitude diagram.}
\end{figure}

\begin{figure}
\center
\includegraphics[scale=0.37, angle=0]{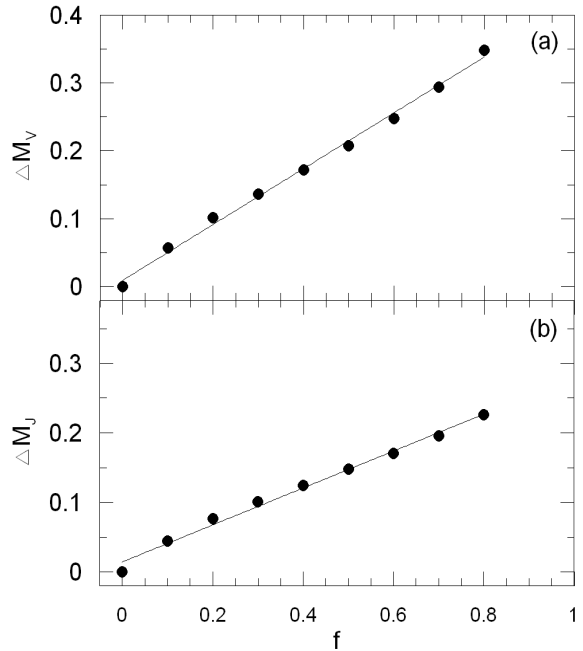}
\caption{Scatter in absolute magnitude versus fraction of binary stars. 
(a) $\Delta M_{V} \times f$ and (b) $\Delta M_{J} \times f$.}
\end{figure}

The last two equations can be used to reduce the estimated absolute magnitudes by 
an amount of scatter corresponding to the adopted fraction of binary stars. The scatter 
$\sigma=0.35$ mag in Table 4 is close to the one of KTG93, i.e. $\sigma=0.27$ mag, 
for the fraction $f=0.8$ of binary stars, confirming our simple but reasonable 
procedure used to reveal the fraction of binary stars and the linear regressions 
in Eqs. (5) and (6).

\subsection {Evolved stars}
A sample of field stars intrinsically brighter than an old star--with the same or 
similar spectral type--passed its turn--off is consisted of evolved stars, and they 
will be brighter and redder than a younger sample. In our case, the contamination 
of the evolved stars seems to be at a minimum due to the restrictions applied to the 
{\em Padova} isochrones in our work, i.e. the sample of the field stars are limited with 
metallicity $-0.30\leq[M/H]\leq0.20$ dex, age $0\leq t\leq 10$ Gyr, and surface 
gravity $\log g>4$. However, we applied the calibrations in Eqs. (3) and (4) to the 
stars of young cluster Hyades and compared the resulting absolute magnitudes with the 
ones estimated by the photometric parallaxes of the cluster stars. After a comprehensive 
study \citet{Perrymanetal98} stated that the Hyades cluster has 282 member stars. However, 
to increase the probability of the membership, we restricted the sample to 141 stars within 
10 pc distance from the cluster center. It turned out that 81 of them were binary 
stars or variable stars \citep[][SIMBAD data center]{Masonetal93, Patienceetal98}. Hence, 
the 60 Hyades stars used in our work are single and non--variable stars within 10 pc from 
the center of Hyades cluster. 

The absolute magnitudes of the Hyades sample estimated by Eqs. (3) and (4) are plotted 
against the ones calculated from their trigonometric parallaxes, taken from the newly 
reduced {\em Hipparcos} data (Fig. 12). There is an agreement between two sets of data, 
indicating that the field sample is not contaminated seriously by the evolved stars. The 
slight declination of the points towards the bright absolute magnitudes is due to single 
epoch observations of the {\em 2MASS} data (Fig. 12b).


\begin{figure}
\center
\includegraphics[scale=0.42, angle=0]{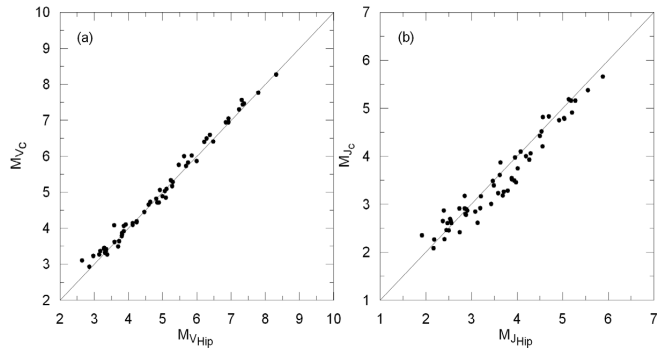}
\caption{Absolute magnitudes, estimated via the calibrations presented in our 
work versus the ones evaluated by the parallaxes in the improved {\em Hipparcos} 
reduction. (a) $M_{V_{c}} \times M_{V_{Hip}}$ and (b) $M_{J_{c}} \times M_{J_{Hip}}$.} 
\end{figure}

We should add that both binarism and evolution effects make the star brighter and redder, 
which means they both make the star move towards the same direction on the colour--magnitude 
diagram. Hence, the scatter in $M_{V}$ and $M_{J}$ absolute magnitudes cited in Eqs. (5) and (6) 
can be assumed as the combined effect of binarism and evolution. In this case, it 
is not surprising that the scatter $\sigma=0.35$ mag cited for the binary fraction 
$f=0.8$ is a bit larger than the one of KTG93, i.e. $\sigma=0.27$ mag, which corresponds 
to the binarism effect alone.

We wish to add that studying the effect of the evolved stars in the Hyades cluster revealed 
an unexpected issue for the cluster. Using the newly reduced {\em Hipparcos} 
data we derived a new distance modulus for the cluster: $3.33\pm0.02$ mag. This value 
is slightly larger than the $3.30\pm0.04$ mag of \citet{Perrymanetal98}.

\section{Acknowledgments}
We would like to thank the anonymous referee for his suggestion towards improving the manuscript. 
We acknowledge the use of the SIMBAD, database, the VizieR Catalogue Service operated at the CDS, 
and the use of the {\em 2MASS} All-Sky Survey. One of us (S.K.) thanks to the Beykent University 
for financial support.

\end {document}